# Magnetic sub-micron rods for quantitative viscosity imaging using heterodyne holography

**C. GENTNER[1], J.-F. BERRET[2], P. BERTO[1,3], S. REICHMAN[1], R. KUSZELEWICZ[1], G.TESSIER[1*]**
[1] Sorbonne Université, INSERM, CNRS, Institut de la Vision, 17 rue Moreau, F-75012, Paris, France
[2] Université Paris Cité, CNRS, Matière et systèmes complexes, 75013 Paris, France
[3] Université de Paris Cité, 45 Rue des saints Pères, F-75006 Paris, France
* gilles.tessier@sorbonne-universite.fr

**Abstract**
Many processes in microfluidics and biology are driven or affected by viscosity. While several methods are able to measure this parameter globally, very few can provide high resolution viscosity images. Optimizing the locality of viscosity measurements demands smaller probes but also shorter lateral diffusion lengths and measurement times. Here, we propose to use sub-micrometer magnetic rods to perform high resolution viscosity imaging. An external magnetic field forces the oscillation of superparamagnetic iron oxide rods. Under linearly polarized illumination, the rotation of these highly anisotropic optical scatterers induces a blinking which is analyzed by heterodyne holography. The spectral analysis of the rotation dynamics yields a regime transition frequency from which the local viscosity is deduced. Holography provides a 3D optical field reconstruction and 3D superlocalization of the rods, which allows super-resolved viscosity measurements. Relying on the fast Brownian rotation instead of the slower translation component of nanorods therefore allows faster measurements and, crucially, smaller effective voxels for viscosity determination. We thus demonstrate that viscosity imaging is possible with a 0.5 $\mu m^3$ 3D resolution.

**Introduction**
Phenomena driven or induced by viscosity are ubiquitous in current biological research [1]. Viscosity has been identified as a key characteristic of cancer cells [2], but is also central to e.g. embryonic development, through the off-center positioning of the oocyte nucleus during meiosis [2, 3]. While macroscopic and global viscosity measurement techniques are well established [4,5,6,7,8,9,10], smaller scale mapping methods are needed to unveil cytoplasm and cytoskeleton structures and behaviors. Micro- or nanoscopic probes are the most common way to obtain small-scale viscosity mapping in a closed environment such as a cell [11]. Viscosity essentially quantifies the response of a medium to mechanical actuation [12]. As such, its measurement requires transducers allowing to first induce a mechanical stimulation, and then detect and analyze the resulting movements in order to deduce mechanical properties. Detection most often relies on optics at the microscopic scale, but several types of actuation methods have been proposed. They can be classified as either *passive* when Brownian motion is used as an inherent mechanical actuation, or *active* when an external force, usually magnetic, is applied to the probe.
Among the *passive methods*, nanoparticle probes are an interesting tool for viscosity measurements. With metallic particles, plasmonic resonances can be exploited to obtain relatively strong scattering cross sections. Illuminating at the appropriate wavelength (plasmonic resonance) and power (to mitigate thermal effects) allows the detection of particles between a few tens and a few hundreds of nm within tens of milliseconds. If the geometry of the particles is well known, the statistical analysis of their Brownian translation is an efficient way to retrieve viscosity [13]. However, the retrieved viscosity values are then averaged along the path followed by the particle during the measurement, which severely limits the spatial resolution. Measuring the Brownian rotation rather than the translation offers an interesting way to increase the locality of the measurement. Fluorescence Polarization Anisotropy techniques measure the polarization of fluorescence emission by freely moving molecules. A polarization change with respect to the excitation reveals molecule rotation and its dynamics, thus providing a measurement of viscosity [14, 15]. Still, the short time scales involved in single molecule rotation and translation are hardly compatible with superlocalization, considering their photon yield. Importantly, mechanical properties strongly depend on the spatial scale of the measurement, and are therefore tightly associated to the size of the probe. The viscosity experienced by nm-range molecules is not invariably relevant at the µm-range scales involved in e.g. intracellular organelles movements in a crowded cytosol.

The optical scattering of non-spherical objects like nanorods is characterized by an anisotropic radiation pattern [16]. As the rod rotates, this results in a strong orientation-dependent modulation of the scattered light [17], which has an optimal contrast when the rod is illuminated near the plasmonic resonance of one of its axes. Nanorods illuminated by a linearly polarized laser beam as they undergo Brownian rotation in a fluid therefore induce a blinking. A spectral analysis of this blinking can lead to local viscosity values, allowing a quantitative viscosity mapping [18]. However, the random nature of the rotation, and of the associated blinking, calls for a careful averaging and an analysis of the frequency spectrum of the optical fluctuations, which is time-consuming and hardly compatible with fast imaging. Fluorescence-based probes can be one order of magnitude smaller than metallic nanorods and are therefore widely used for microrheology[19,20,21]. Rather than translation, they rely on fluorescence lifetime imaging (FLIM) and on the rotation of the probe molecules which also occurs on shorter time scales than translation processes [10, 11, 12,22,23]. These passive methods, however, rely on the statistical analysis of random Brownian fluctuations and therefore demand the measurement of many events, i.e. relatively long acquisitions, to obtain significant results.

For this reason, *active methods* use a deterministic actuation of the probe movement. They can increase the amplitude and efficiency of the mechanical actuation, but also authorize faster measurement by reducing the need for statistical analysis and by allowing the use of signal processing strategies. Atomic Force Microscopy being mostly limited to surface measurements [24], optical tweezers are an interesting tool to induce movement inside the medium. They can drive particles with high precision, and allow measurement durations as short as tens of µs [25,26,27,28]. Spatially resolved measurements or even imaging can then be obtained using multiple optical tweezers and/or particle scanning, but these methods are relatively complex and cumbersome. External magnetic actuation is probably the preeminent method to induce probe movements, as it is contactless and offers excellent penetration depths in most media. In several microrheology experiments [12, 22, 23, 29, 30, 31], maghemite rods (a polymorphic iron oxide [31, 32]) are rotated using various revolving magnetic field frequencies. Determining the cut-off frequency at which rods cease to follow the actuation provides an accurate measurement of the viscosity. However, since rotation is detected using direct optical imaging, the rods must be resolved, and their sizes typically lie in the 10 µm range. While intracellular measurements have been reported, they can hardly be considered local since the probe size is comparable to that of most cells [31].

Improving the spatial resolution of viscosity measurements of active and passive methods is essential, particularly for biological applications: subcellular measurements require sub-10 μm spatial resolution. To obtain high resolution viscosity measurements, a small probe size $d$ is of course essential. However, during the measurement time $\tau_{meas}$, untethered probes undergo a lateral Brownian diffusion driven by their translational diffusion coefficient $D_t$. The viscosity measurement is therefore averaged over the diffusion length $L = \sqrt{D_t \tau_{meas}}$ by which the probe travels while viscosity is being measured. For this reason, the spatial resolution of the viscosity measurement is at best limited by the largest of these parameters, $Max\left(d, \sqrt{D_t \tau_{meas}}\right)$. In most cases, the diffusion-length term dominates over $d$ and, rather counterintuitively, a smaller probe size $d$ can lead to a degraded resolution. Indeed, small probes have larger diffusion lengths $L$ due to both a larger diffusion coefficient ($D_t \propto \frac{1}{d}$) and smaller optical yields which call for longer detection times $\tau_{meas}$. For this reason, a viscosity measurement with optimal spatial resolution requires the simultaneous minimization of all three parameters, $d$, $D_t$, and $\tau_{meas}$.

In this paper, we report on a fluorescence-free active microrheological method using an external magnetic actuation to rotate superparamagnetic rods, aiming at resolution improvement by playing on each of these parameters: *i)* The size of the probes $d$ is decreased to the micrometer, i.e. one order of smaller than existing methods [12, 22, 23, 29, 30, 31]. *ii)* By relying on the fast rotation of nanorods rather than on a slower translation, lateral displacement during the measurement is reduced ($D_t < D_r$). Finally, *iii)* a magnetic actuation induces deterministic movements at a chosen frequency which is selectively tuned to the locked-in heterodyne frequency of a shot noise-limited holographic setup, thus improving detection efficiency and reducing the acquisition time $\tau_{meas}$ for small particle detection. We use maghemite magnetic rods with sizes near the optical diffraction limit. Direct imaging cannot provide an accurate measurement of the orientation of such rods, but the strong anisotropy of their effective scattering cross section induces a modulation of light scattering at the rotation frequency (or its harmonics), i.e. a blinking. The use of relatively high-frequency (kHz) magnetic field modulations, enabled by the small size of the probes ensures an efficient rejection of noises by lock-in detection. A heterodyne holographic system allows a frequency-selective multiplexed lock-in detection at the modulation frequency, rejecting noises and Brownian contributions at other frequencies. In addition, dark field illumination and the holographic gain allowed by the reference beam ensures a sensitive, shot-noise-limited detection, thus optimizing the detection times. Using dense rod seeding, we show that this method provides spatially-resolved images of viscosity gradients with optical resolution. However, with single, isolated rods, digital holography allows the reconstruction of 3D images of light scattering, and therefore a 3D superlocalization of rods [33]. We show that a measurement of the viscosity can be obtained with a single rod within $\tau_{meas} = 2$ s, a time during which the stochastic exploration of the rod is limited to a 0.5 μm$^3$ volume, of the same order as the hydrodynamic volume of the rod, thus paving the way for high resolution, 3D viscosity mapping.

1. **Optical Method**

When aiming at higher spatial resolution, the size of the probes is a key issue. Since suitably small magnetic nanorods are not commercially available, sub-micrometer maghemite rods (length $L = 1020 \pm 240$ nm, diameter $D = 440 \pm 80$ nm, mean values and standard deviations measured by imaging 50 rods magnetically aligned in the image plane) were synthesized (Supplementary Information 1). As shown in Fig.1a, the orientation of these rods is driven by an external magnetic field modulated at a frequency $f_B$, superimposed to a perpendicular static field, thus creating an oscillating effective magnetic field $\vec{B}(f_B)$.

The optical detection of the rods is based on the blinking resulting from rod rotation as described previously for nonmagnetic objects [34]. The scattering diagram of a nanorod is essentially a torus aligned with the long rod axis, as schematically depicted in Fig.1a. When illuminated by a 785 nm laser beam with a polarization $\vec{E_O}$, the intensity and angular distribution of the electric field $\vec{E_S}$ scattered by a rod depend on its orientation with respect to $\vec{E_O}$. The portion of this light collected by a microscope objective is therefore modulated as the rod rotates. A Digital Holographic Microscope (DHM) operating in homodyne or heterodyne lock-in mode [35] selectively measures the amplitude of the collected light modulated at a chosen frequency $f$ (see section 2). This interferometric imaging method provides two main advantages. First, DHM exploits the product $\vec{E_S} \cdot \vec{E_R}$ of the electric field $\vec{E_S}$ scattered by the object (i.e. the light scattered by a nanorod) by a reference field $\vec{E_R}$. Although $E_S$ is weak in the case of nanorods with small scattering cross section, $E_R$ can be easily increased to obtain hologram intensities approaching the saturation level of the camera, thus providing an optimal, shot noise-limited detection of small nanoparticles [36, 34, 28, 19]. Second, the two-arm interferometric configuration (described later in fig. 2a) is particularly suited to single- (homodyne) or dual- (heterodyne) frequency phase modulations. As such, it allows the study of modulated phenomena over a very broad frequency range (from mHz to tens of MHz) [35,36 ,37, 38], and thus gives access to high rotation/blinking frequencies.

As shown in Fig.1b, when monitoring a single rod, the presence of a magnetic field $\vec{B}$ modulated at $f_B = 10$ Hz induces an increase in the DHM signal at the frequency $f = f_B$ which clearly indicates that the rod rotates and blinks at the frequency $f_B$. However, as shown in the inset, the object does not disappear totally in the absence of modulation: Brownian motion randomly contributes to a component at $f$.

A simultaneous scanning of the magnetic field and analysis frequencies $f_B$ and $f$ yields a spectrum which displays two distinct frequency regimes separated by a transition at a critical frequency $f_t$, as shown in Fig.1c (see Supplementary Information 2 for a complete derivation):

- $f \ll f_t$ : the magnetic field oscillation period is longer than the characteristic time taken by the magnetic moment to align. The rod is able to follow the rotation of the magnetic field, and the signal amplitude is then maximum.
- $f \gg f_t$ : the magnetic field oscillation period is too short to allow the rod to align with it. The average orientation of the rod long axis is thus stuck in the direction of the permanent magnetic field $\vec{B_0}$, and the amplitude of the modulated component of the DHM signal decreases to zero as $f_B$ increases.

This transition frequency $f_t$ is clearly related to viscosity $\eta$. Its measurement therefore gives access to $\eta$ through Eq. (1), which is obtained by considering the evolution of a superparamagnetic rod, i.e. a *macro*-magnetic moment $\vec{m} = \frac{\chi V}{\mu_0}\vec{B}$ [18], in a field $\vec{B} = B_\sim \cos(2\pi F_B t)\vec{u_z} + B_0\vec{u_x}$. (See SI 2).

$$f_t = \frac{1}{2\pi\mu_0}\frac{\Delta\chi B^2}{\eta} h\left(\frac{L}{D}\right) \quad (1)$$

where $\eta$ is the medium dynamic viscosity, $h$ is a function of the rod length $L$ and diameter $D$ defined as $h\left(\frac{L}{D}\right) = \frac{L^2 N_\parallel + D^2 N_\perp}{2(L^2+D^2)}$ for a prolate spheroid. Here, the spheroidal depolarization factors can be expressed as $N_\parallel = \frac{1-e^2}{2e^3}\left(\ln\left(\frac{1+e}{1-e}\right) - 2e\right)$, $N_\perp = \frac{1-N_\parallel}{2}$, with an eccentricity written as $e = \sqrt{1-\left(\frac{D}{L}\right)^2}$. The magnetic susceptibility anisotropy factor for a rod composed of an assembly of spherical nanoparticles of bulk susceptibility $\chi$ can be written $\Delta\chi = \frac{\chi^2(N_\perp - N_\parallel)}{(1+N_\parallel\chi)(1+N_\perp\chi)}$ [39]. The factor $B^2$ in Eq. (1) corresponds to the squared modulus of the average magnetic field, $B^2 = B_0^2 + \frac{B_\sim^2}{2}$.

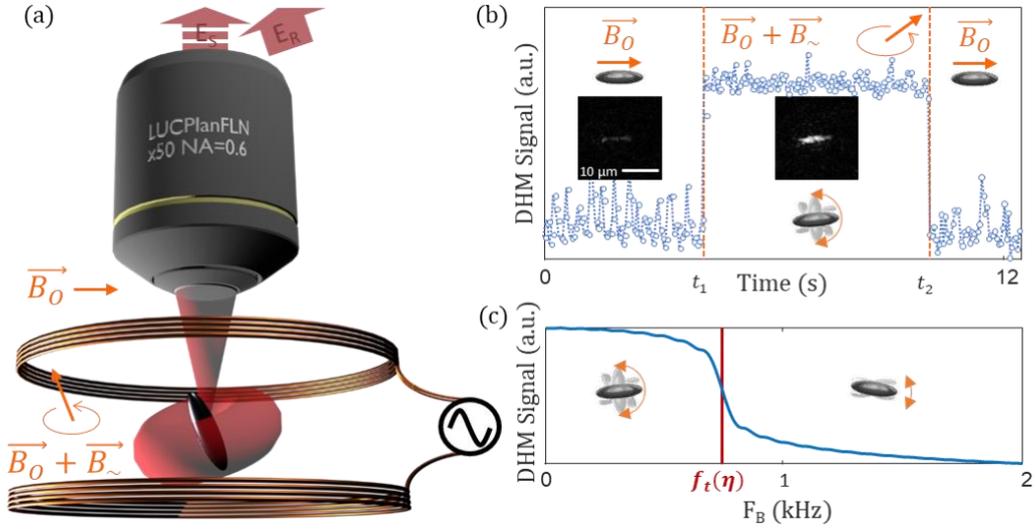

**Figure 1:** (a) Principle of the measurement: the orientation of a maghemite nanorod is driven by the superposition of a static magnetic field $\vec{B_0}$ (horizontal) created by a Nd$_2$Fe$_{14}$B permanent magnet, and an oscillating field $\vec{B_\sim}$ (vertical) generated by coils and modulated at a frequency $f_B$. Part of the torus-shaped optical scattering diagram is collected by the microscope objective and sent to a heterodyne Digital Holography Microscope (DHM). (b) Reconstructed DHM images (insets) and average signals acquired at an analysis frequency $f = f_B = 10\ Hz$. When $\vec{B} = \vec{B_0}$ ($\vec{B_\sim} = 0$ for $t \in [0, t_1[$), the rod is mostly static, aligned with $\vec{B_0}$, and the $f_B$ component of the DHM signal is low. When $\vec{B} = \vec{B_0}+\vec{B_\sim}$ ($t \in [t_1, t_2[$), the strong DHM signal at $f_B$ indicates that the rod oscillates and blinks at $f_B$. (c) Numerical model of the DHM signal, a sigmoid with a *sinc* acquisition transfer function [18].

As shown in S.I. 2, the measured spectrum can be modeled as a sigmoidal function convolved with a *sinc* transfer function [18]. The inflexion point of this sigmoid indicates the transition frequency $f_t$. For a rod of dimensions $L$ and $D$ in an oscillating magnetic field of amplitude $B$, a fit of this spectrum using Eq. (1) with $f_t$ as parameter can therefore provide a local value of the viscosity $\eta$ across the volume explored by the rod during the acquisition.

The expression of $f_t$ in Eq. (1) is similar to that of the more widely studied case of a rotating field [22, 30, 31], although the transition is broader: as demonstrated in Supplementary 2, when the oscillating field frequency increases, the angular exploration range of the rod decreases smoothly from its maximum (lower than $\pi$) to 0 within the same half-plane; whereas a rotating field produces a sharp cutoff between a full $2\pi$ rotation regime and a regime with oscillation around a fixed (arbitrary) axis. Rotating fields are typically produced using four identical coils [12, 22, 23, 31]. Here, we use a dark field illumination through an oil-coupled prism (not shown in Fig.1) to avoid saturating the camera with the illumination beam. This prism strongly constrains the available space around the sample, and we chose a more compact oscillating field configuration.

The frequency range of the transition frequency $f_t$ is driven by the viscosity value. For high viscosities, the frequencies can be low enough to be sampled by a camera, in a homodyne configuration (section 2 below). However, for lower viscosities, $f_t$ can become too high (typ. above a few tens of Hz) to be properly sampled by a camera, and we propose the use of a heterodyne holographic detection, as detailed in section 3.

## 2. Viscosity measurement via homodyne holography

The first experiment, shown in Fig.2, was carried out in two microfluidic chambers separated by a horizontal 300 µm PDMS wall. The first chamber (Fig.2b-top) contains a 27% glycerol aqueous solution ($\eta = 2.6$ mPa s). The second chamber contains water ($\eta = 0.89$ mPa. s) for comparison. The $1020 \times 440$ nm maghemite nanorods were synthesized using a low weight concentration ($c = 0.02\ wt\%$) of initial nanoparticle-polymer solution using a synthesis protocol described in [40, 41]. Individual nanorod detection requires a density well below the limit of 1 object per optical Point Spread Function (PSF), i.e. a few $10^6$ rods.µL$^{-1}$ with the x4, NA 0.13 objective we used. Here, the rods were diluted to reach a concentration $10^3$µL$^{-1}$ in both chambers, therefore 3 orders of magnitude below this limit, ensuring that less than one nanorod is statistically present within each PSF. The rods were submitted to an oscillating magnetic field $\vec{B} = \vec{B_0} + \vec{B_\sim}$. The static component $\|\vec{B_0}\| \approx 1\ mT$ is generated using a Nd$_2$Fe$_{14}$B permanent magnet and measured with a Hall probe (Leybold). While the amplitude of the modulated component $\|\vec{B_\sim}\|$ could not be measured directly due to the relatively long response time of this probe, $B_\sim = 1.3 \pm 0.3$ mT was estimated by feeding coils with a DC current equal to the amplitude of the intensity modulation..

The holographic detection setup is depicted in Fig.2a. To allow an intense illumination of the low-scattering-cross-section nanoparticles without blinding the camera, the beam illuminating the microfluidic chamber is sent through a glass prism coupled to the microfluidic chamber, and totally reflected at the glass-air interface. Only the light scattered by the nanorods is collected by a long working distance, X4 magnification and NA=0.13 microscope objective, and sent to the camera. There, it interferes with a reference plane-wave coming from the same 785 nm, single longitudinal mode laser in an off-axis configuration, at ~1° tilt angle. To extract the component of the signal modulated at $f_B$, the camera is triggered at $f_{acq} = 4f_B$, in a 4-bucket integration scheme [37, 41] which filters out fluctuations occurring at other frequencies and selectively produces holograms of the $f_B$-modulated component. After a first Fourier transform, the +1 interferometric order is numerically extracted, propagated, and finally back Fourier-transformed to obtain the image of a chosen plane inside the microfluidic chamber [39].

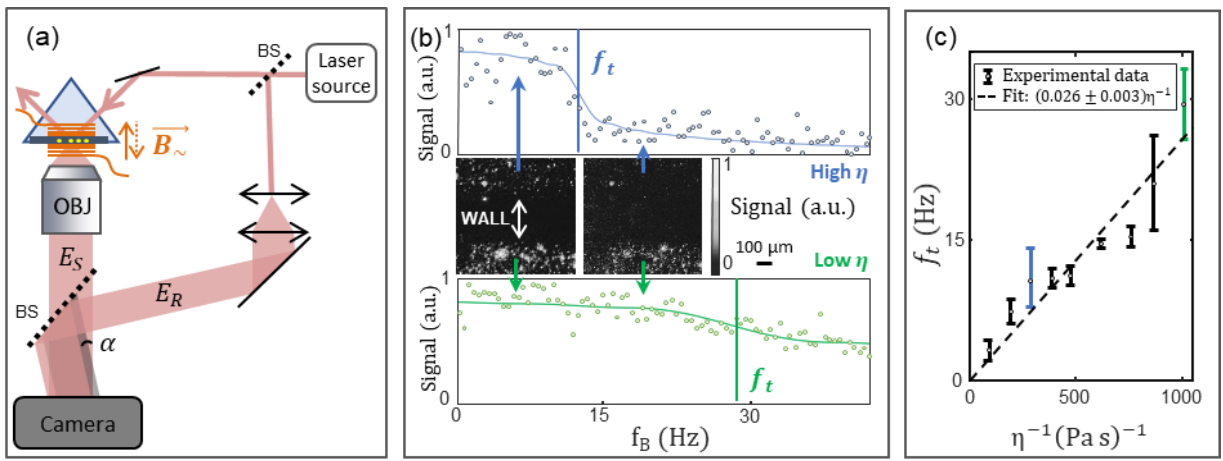

*Figure 2:* (a) Holographic setup. A magnetic field oscillating at $f_B$ modulates the orientation and optical scattering of rods placed in a microfluidic chamber illuminated in total internal reflection. The holograms are acquired in an off-axis configuration with an angle ⍺ between the object and reference beams, sampled at $f_{acq} = 4f_B$, demodulated and reconstructed. (b) Images of two chambers containing water-glycerol mixtures of viscosity $2.6 \, mPa \, s$ (top) and $1.0 \, mPa \, s$ (bottom) separated by a horizontal wall, at frequencies $f_B = 7 \, Hz$ and $f_B = 20 \, Hz$. Spectra of signal averaged in the high viscosity (top, green) and low viscosity regions (bottom, blue) exhibit different transition frequencies $f_t$, at $12 \, Hz$ for $\eta = 2.6 \, mPa \, s$ and $29 \, Hz$ for $\eta = 1.0 \, mPa \, s$. (c) Transition frequency $f_t$ measured in solutions of 0 to 55% water-glycerol mixtures, plotted against the expected inverse viscosity $\eta^{-1}$ estimated using the known glycerol concentration [45]. The slope of the linear fit (0.026 Pa, dashed line) is in good agreement with the slope deduced from Eq.1 (0.030 Pa). The values of $f_t$ deduced from the fits shown in b) are plotted in blue and green. Error bars : dispersion obtained on 10 different rods.

For each frequency $f$ in the range [0.5,42] Hz, series of 80 holograms were acquired, yielding 20 holograms after 4-bucket demodulation. Fig.2b shows images reconstructed from the average of these 20 holograms obtained at 7 and 20 Hz. Similar images and signals averaged in each compartment were acquired in 0.5 Hz steps. The values of $f_t$ were extracted from fits using the function derived in detail in the Supplementary section. Note that a simplified sigmoidal function was used here, omitting the convolution by the $sinc$ function for sake of computational speed.

In the high viscosity compartment ($\eta = 2.6$ mPa.s, green curve in Fig.2b), a clear transition frequency $f_t = 12 \pm 1$ Hz is observed. With $\frac{L}{D} = 2.8 \pm 0.5$ and $\chi = 6.5$ [31], Eq. (1) predicts $f_t = 12 \pm 7$ Hz for $B_0 = 0.9 \pm 0.3$ mT, in good agreement with the measured value. In the lower viscosity medium, $\eta = 1.0$ mPa s, we measure transitions at $f_t = 29 \pm 5$ Hz, which compares well with the expected value $f_t = 32 \pm 22$ Hz. Images acquired between these two values of $f_t$ (Fig.2b, center, $f = 20$ Hz) clearly display a viscosity contrast: while rods do not follow the rotation of the magnetic field in the high viscosity medium (top), they are still rotating and appear bright in the low-$\eta$ region (bottom). Similar spectra were acquired to measure $f_t$ in various water/glycerol mixtures, with glycerol concentrations in the [0%, 55%] range, corresponding to viscosities in the [1, 10] mPa.s range (deduced using [42]). The measured values $f_t(\eta^{-1})$ plotted in Fig.2c display a linear behavior, with a best fit slope of $0.026 \pm 0.003$ Pa. This is in good agreement with the value $f_t \eta = \frac{\Delta \chi B^2 h\left(\frac{L}{D}\right)}{2\pi \mu_0} = 0.03 \pm 0.01$ Pa predicted by Eq. (1) with the parameters used in our experiment, $\Delta \chi = 2.0 \pm 0.1$, $\frac{L}{D} = 2.8 \pm 0.5$, $B_0 \approx 1$ mT and $B_\sim = 1.3 \pm 0.2$ mT. These results show that the measurement of $f_t$ can be used to deduce the viscosity $\eta$ using either equation (1) if all the necessary parameters are known, or a calibration curve based on measurements of $f_t$ in solutions of known viscosity, as shown in Fig. 2c.

However, the accurate determination of the transition frequency requires measurements at frequencies $\frac{f_{acq}}{4}$ well above the value of $f_t$. With $f_t = \frac{\Delta \chi B^2 h\left(\frac{L}{D}\right)}{2\pi \eta \mu_0}$, this can be difficult to achieve with most cameras ($f_{acq}$ of a few tens of Hz), particularly at low viscosity values.

A possible workaround is to decrease the intensity of the modulated component of the magnetic field $B_\sim$. However, this approach breaks down when the magnetic driving force becomes of the same order as the random Brownian forces exerted on the particle. Before this, lowering $B_\sim$ clearly decreases the detection signal-to-noise ratio. If we choose an *arbitrary* limit of $10^3$ for the ratio between the magnetic energy $mB$ and the energy of Brownian fluctuations $k_B T$ which allows a satisfactory detection, i.e. $\frac{mB}{k_B T} > 10^3$, we can evaluate the minimal magnetic modulation $B$ which is required. Superparamagnetic rods have a macroscopic magnetic moment amplitude $m = \frac{\chi V}{1+N_\parallel \chi} \frac{B}{\mu_0}$ where $V$ is the rod volume. For our rod size and composition, the minimum magnetic field amplitude needed to overcome Brownian fluctuations is then $B = 3.5$ mT, which, from Eq. (1), corresponds to a transition frequency $f_t = 200$ Hz in water. In 4-bucket integration mode, its measurement therefore requires camera frame rates well above $4f_t = 800$ Hz. Cameras operating in the kHz range or above are available, but sub-ms integration times are hardly compatible with nano-object detection. For this reason, we have chosen to use heterodyne holography, which allows the measurement of high frequency phenomena using slow detectors [18, 35, 43, 36, 37, 38, 44 , 45, 46].

### 3. Viscosity mapping via heterodyne holography

Heterodyne detection is implemented by adding an acousto-optic modulator (AOM) in each holographic path, as illustrated in Fig.3a, to introduce a relative phase modulation at a chosen frequency. Since AOMs operate in the 80 MHz range, the object beam is modulated at 80 MHz, while the reference beam is modulated at 80 MHz $- \Delta f$ to produce an optical phase beating at $\Delta f$. The 4-bucket integration described above is used to sample and demodulate holograms with a camera triggered at $f_{acq}$. By choosing the beating frequency so that $\Delta f = \frac{f_{acq}}{4} + f_i$, one can selectively isolate holograms of any modulation at $f_i$. Much like a multiplexed lock-in scheme, this process isolates a hologram of component modulated at frequency $f_0$ and filters out other fluctuations (including most of the Brownian noise in our case). Even with a slow camera, i.e. $f_{acq}$ of a few Hz, heterodyning provides high-Signal-to-Noise-Ratio (SNR) -holograms of phenomena modulated at frequencies $f_i$ up to tens of Hz.To illustrate the ability of this system to image quantitatively the viscosity, a test sample displaying a spatial viscosity gradient was prepared using a concentration $c_{Agar} = 0.5\ wt\%$ Agar gel ($\eta \approx 5$ mPa s ▯▯▯▯) and water ($\eta = 0.89$ mPa s). These media (agar and water) were put in contact to obtain a gradient of Agar concentration by diffusion at the interface. In order to visualize this gradient, 1% Alexa 488 fluorophores were added to the water phase. The fluorescence image shown in Fig.3c is therefore directly related to the water gradient ($1 - c_{Agar}$), with a higher agar concentration and higher viscosity on the right, and a lower agar but higher Alexa fluorophore concentration on the left of the image. Both media were homogeneously seeded using the same maghemite nanorods ($L = 1020 \pm 240$ nm, $D = 440 \pm 80$ nm) at a concentration of $10^5 \mu L^{-1}$. With the microscope objective used here (×4 ON 0.13), this corresponds to about 100 rods per PSF, i.e. rods were not individually resolved.

In the experiment shown in Fig.3, the coils were powered using a 100 V – 200 W audio amplifier to produce a modulated field amplitude $B = 3.9$ mT. This stronger field modulation (as compared to B = 1.3 mT in the previous experiment) increases SNR, but also leads to higher transition frequencies $f_t$. Heterodyning, however, gives access to a broad frequency range. For each of the 27 measured frequencies $f_i$ (i.e. $i = 1$ to 27) ranging from 80 to 1000 Hz, 100 holograms were acquired at $f_{acq} = 10$ Hz, with a $\tau = 50$ ms integration time. After 4-bucket demodulation of the 100 measurements, the 25 resulting holograms were averaged and numerically propagated to the sample plane [47] to produce 27 images of the dynamic blinking for each of the 27 frequencies $f_i$. The $1024 \times 1024$ pixels image were then resampled into 12 x 12 macropixels, a size corresponding to $65 \times 65\ \mu m^2$ in the object space, which was chosen as a trade-off between spatial resolution and SNR. The resulting hyperspectral (12 x 12 x 27) data was then used to obtain spectral fits (i.e. along $f_i$) using the sigmoidal fitting function described above (see Supplementary section for the derivation of this function). Note that the slope parameter of the sigmoid was set equal to the value of the camera integration time as being the slowest response time of the system. One of these fits is shown in Fig. 3b, and corresponds to one of the 144 pixels (indicated by a dashed box in Fig. 3c). These fits yield transition frequencies $f_t(x, y)$ for each pixel, which are converted into a 12x12 image of the local viscosity values $\eta(x, y)$ using Eq.(1), as shown in Fig. 3c.

As shown in Fig. 3c, the fluorescence image indicates that a horizontal viscosity profile is expected, with quasi-pure water on the left, and a low Agar concentration (< 5%) to the right of the image. This is indeed observed qualitatively in the viscosity image (Fig.3 d), although some evolution of the gradient is likely to occur during the 5 min total acquisition time. The [245,275] Hz range of $f_t$ values corresponds to a viscosity in the [1.1,1.3] mPa s range, close to values expected in pure water, i.e. very low Agar concentration. This underlines the sensitivity of the method, which allows to identify weak viscosity variations.

### 4. Reducing the voxel size by single particle tracking

The experiments shown in Fig.3 involve relatively large macropixels (65 µm laterally), each addressing a large number (typ. thousands) of nanorods. With sufficient SNR, this volume can be easily reduced down to the size of the PSF of the microscope. Further reduction is possible provided that less than one particle is present in the volume of the PSF: calculating the center of mass of the holographic image of a rod provides its 3D position with nm-range precision. In liquids, Brownian motion ensures a stochastic exploration of the environment by the rods, and we have shown that this can provide sub-diffraction optical imaging in 3D [48]. As discussed above, the Brownian rod is free to travel laterally over a mean distance $\sqrt{D_t \tau_{meas}}$ during the measurement time $\tau_{meas}$ ($D_t$ is the lateral diffusion coefficient), which determines the size of the probed volume, i.e. the spatial resolution.

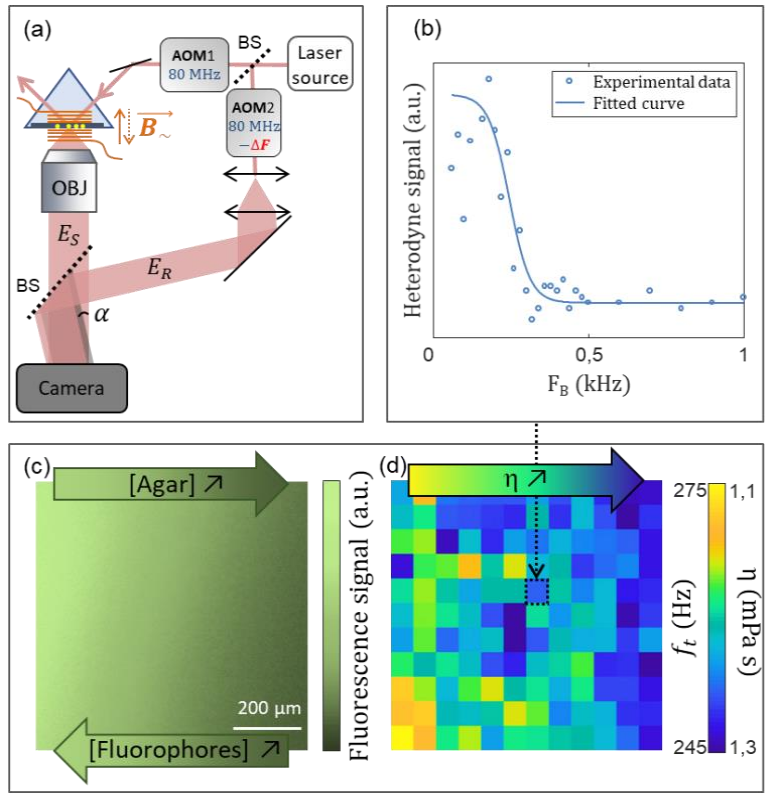

*Figure 3*(a) Heterodyne holography setup: acousto-optic modulators (AOM) are added in both the object and reference arms of the holographic setup to induce a beating at $\Delta f = \frac{f_{acq}}{4} + f_i$ and selectively selective retrieve $f_i$ components after 4-phase demodulation. (b) Experimental spectrum and sigmoidal fit obtained for one of the pixels of (d), indicated by the dashed arrow. (c) A sample displaying an agar concentration gradient was prepared by diffusion of Agar (left) into water containing Alexa fluorophores (right). Fits similar to b) were conducted on each (x,y) pixel, yielding (d) an image of the local transition frequency $f_t$ or of the quantitative viscosity. The measured viscosity (d) increases from left to right. Note the opposite color scales for $f_t$ and $\eta$ since $f_t \propto \eta^{-1}$: higher $\eta$ values are blue (right of the image) while low viscosities are yellow (left of the image), in agreement with the qualitative fluorescence and agar gradients.

Fig. 4a shows the trajectories of a nanorod measured by superlocalization on holographic reconstructions recorded at $f_{acq} = 20$ Hz with a ×50, 0.6 NA objective, with a magnetic field $B_\sim = 3.5 \pm 0.3$ mT and $B_0 = 3.0 \pm 0.4$ mT. The volume explored by the rod during $\tau_{meas} = N/f_{acq}$ (with $N = 40$ to 1000 the number of holograms) decreases down to 0.5 µm³ for $N = 40$. Clearly, the SNR of the heterodyne signal also decreases with $N$. However, the transition frequency $f_t$ is well identified in measurements obtained at 9 frequencies by averaging $N = 40$ holograms in 3 glycerol solutions of respective viscosities $1.0 \pm 0.1$, $2.1 \pm 0.2$ and $3.0 \pm 0.4$ mPa s. Fits on these data yield viscosities in good agreement with these expected values, showing that $N = 40$ averages provide a reliable measurement of the viscosity, as shown in Fig. 4b). Several frequencies $f_i$ are arguably necessary to derive these values, but assuming that only one measurement is taken, which would be possible by heterodyne frequency multiplexing [49] [44], the value $(0.5 \text{ µm}^3)^{1/3} = 800$ nm therefore indicates the best achievable resolution at this point. This corresponds to the diffraction limit $\left(\frac{1.22\lambda}{2NA} = 800 \text{ nm}\right)$ laterally, but clearly surpasses it longitudinally. With higher SNR, by e.g. increasing the optical intensity or $B$, optical superresolution by 3D rod superlocalization is therefore achievable.

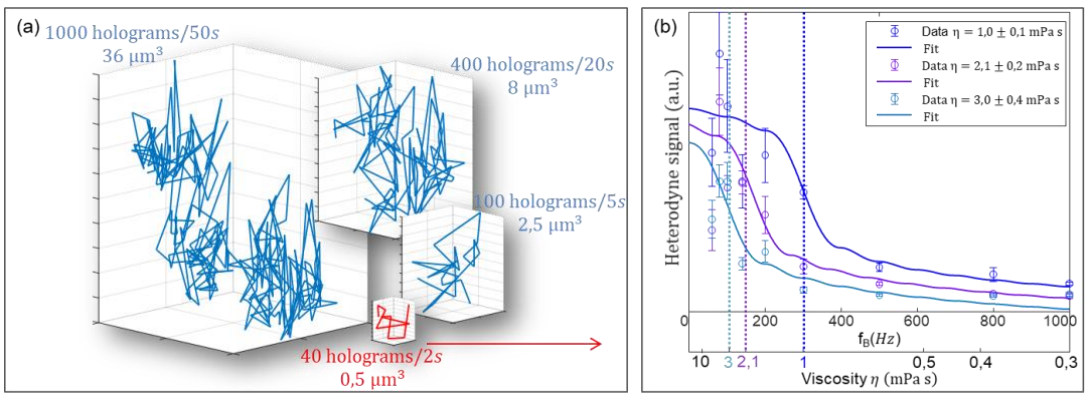

Figure 4: (a) Volumes explored by a maghemite nanorod during $N = 40$ to 1000 heterodyne holography measurements at a rate $f_{acq} = 20 Hz$. The trajectories were determined by superlocalization on 3D reconstructions. (b) Spectra of single magnetic rod rotation in water-glycerol mixtures with viscosities $1.0 \pm 0.1$, $2.1 \pm 0.2$ and $3.0 \pm 0.4$ mPa s with $B_\sim = 3.5\ mT$. The fits are performed by convolution between the $sinc$ instrumental transfer function and a sigmoid function (see S.I), with two free parameters: the transition frequency $f_t$ and an offset $a$.

**Conclusion**

Although well tolerated and easily internalized by living cells, optically resolved maghemite needles with lengths of the order of 10 μm are not adapted to sub-cellular resolution measurements. Passive, Brownian-motion based methods with sub-100 nm gold nanorods [18] are a possible alternative since they are also biocompatible, and offer relatively large optical scattering cross sections near their plasmon resonance. However, analyzing random motion to extract viscosity values demands a fine sampling and long acquisition times which reduce the effective spatial resolution of the viscosity image. The sensitivity of the holographic detection proposed here definitely allows a drastic reduction of the probe size compared to the works which inspired it [12, 22, 23, 31]. We have shown here that the magnetically-driven rotation of non-plasmonic rods can be detected selectively at frequencies up to the MHz for $L \approx 1$ μm. The current photonic budget of the holographic detection indicates that a size reduction down to a few hundreds of nm is possible, and could be further reduced by increasing the scattering cross section by e.g. gold-coating of the maghemite rods.

To further improve the locality and the spatial resolution, fast- and high-SNR-measurements are crucial. Increasing the amplitude of the modulated field $B_\sim$ is an efficient way to overcome rotational Brownian fluctuations. As discussed here, the transition frequency $f_t$ increases with $B_\sim^2$ and this points to high frequency measurements. Up to the kHz, high speed cameras are a solution, but short exposures call for intense illuminations which can affect the local viscosity by heating up the probes. The heterodyning scheme used here can isolate blinking nanorods well above the MHz using mW-range illumination. Since two frequencies are in principle sufficient to determine $f_t$ in media where the expected viscosity range is known, holographic multiplexing is also a possible way to improve measurement speed and hence spatial resolution.

The geometry of the holographic assembly, the vertical direction of the rotation that it imposes and the presence of the prism (for dark-field illumination) complicate the use of a rotating field scheme by four identical coils [12, 22, 23, 31]. The oscillating field configuration shown in figures 2 and 3 is a good compromise between simplicity of fabrication and large magnetic field modulation. However, oscillating fields induce a smoother inflection (see S.I. 2) than the sharp transition observed with fully rotating fields, thus reducing the accuracy of the measurement.

Holography has the advantage of allowing 3D reconstructions. Although the PSF extension is larger in the axial direction, three-dimensional particle localization (Fig.4) can provide 3D viscosity mapping. Conversely, without localization, a pixel contains the average of signals from multiple planes (Fig.3c). There is no axial sectioning, except for the weighting of out-of-focus signals according to their pixel spread. This can be problematic in the case of thick and highly scattering samples. Holography with low coherence length sources is a possible strategy to obtain axial sectioning, in much the same way as Optical Coherence Tomography (OCT).

**Acknowledgements** : The authors wish to thank Marilou Clémençon, Christophe Tourain, Vincent de Sars, Marie-Hélène Verlhac, Maria Almonacid and Noemi Zollo for help in the preparation of various samples and experiments. ANR (Agence Nationale de la Recherche) and CGI (Commissariat à l'Investissement d'Avenir) are gratefully acknowledged for their financial support of this work through ANR-21-CE42-0008 ELISE, ANR-10-LABX-0096 and ANR-18-IDEX-0001, ANR-17-CE09-0017 (AlveolusMimics) and ANR-21-CE19-0058-1 (MucOnChip), Labex SEAM (Science and Engineering for Advanced Materials and devices). We acknowledge the ImagoSeine facility (Jacques Monod Institute, Paris, France), and the France BioImaging infrastructure supported by the French National Research Agency (ANR-10-INBS-04, « Investments for the future »).